**Spin gapless semiconductors**

Zengji Yue[1,2]*, Zhi Li[1,2], Lina Sang[1,2], and Xiaolin Wang[1,2]*

1. Institute for Superconducting and Electronic Materials (ISEM), Australian Institute of Innovative Materials (AIIM), University of Wollongong, North Wollongong, NSW 2522, Australia.

2. ARC Centre of Excellence for Future Low-Energy Electronics Technologies (FLEET), University of Wollongong, North Wollongong, NSW 2522, Australia.

Email: zengji@uow.edu.au; xiaolin@uow.edu.au;

**Abstract**

Spin gapless semiconductors (SGSs) are a new class of zero gap materials which have a fully spin polarised electrons and holes. They bridge zero gap materials and half-metals. The band structures of the SGSs can have two types of energy dispersions: Dirac linear dispersion and parabolic dispersion. The Dirac type SGSs exhibit fully spin polarized Dirac cones, and offer a platform for massless and fully spin polarized spintronics as well as dissipationless edge state via quantum anomalous Hall effect. Due to its fascinating spin and charge states, they hold great potential application in spintronics. There have been tremendous efforts worldwide on searching for suitable candidates of SGSs. In particularly, there is an increasing interest in searching for Dirac type SGSs. In the past decade, a large number of Dirac or parabolic type SGSs have been predicted by density functional theory and some of parabolic SGSs have been experimentally demonstrated. The SGSs hold great potential for high speed and low-energy consumption spintronics, electronics and optoelectronics. Here, we review both Dirac and parabolic types of SGSs in different materials systems and outline the concepts of SGSs, novel spin and charge states, and potential applications of SGSs in next generation spintronic devices.

## 1. Physical properties of SGS materials

1.1 The definition of SGS materials

Generally, all materials can be divided into three types: metals, semiconductors, and insulators based on their electronic band structures. While gapless materials or zero gap materials have unique electronic, photonic and magnetic properties as compared to conventional gapped materials or metals.[1] The discovery of graphene which is a Dirac type zero-gap material has generated great interests in exploring new classes of zero-gap materials. The spin-gapless semiconductors (SGS) was conceptually proposed and coined by Wang (one of the co-authors of this article) in 2008 through the designs of band structures.[2]

There are numbers of unique features for the SGSs.[1,2,3] The conduction band (CB) bottom touches the valence band (VB) top at Fermi level i.e the energy gap is zero and both electrons and/or holes are fully spin polarized. The energy dispersion of the SGSs band structure can be either linear or parabolic. The SGSs with linear energy dispersion was coined as Dirac SGSs in 2017.[3] Novel spin or charge related Hall effect has also been predicted in this article [3], followed by an extended review on Dirac SGSs.[4] The general definition of SGSs is the following: Spin gapless semiconductors (SGSs) are zero-gap materials with zero-gap for one spin channel and a gap for the other spin channel. However, four different types of SGS materials with different configurations of band structures were originally proposed. The four types of either parabolic or Dirac SGSs with different configurations of spin polarization for both conduction bands and valences band are schematically shown in Figures 1and 2. Their corresponding density of states are also illustrated in Figures 1 and 2.

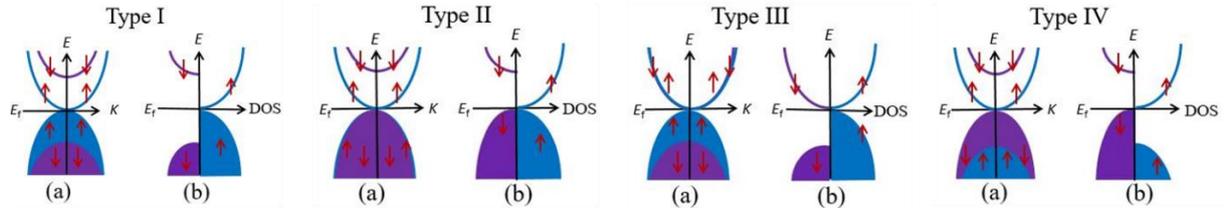

Figure 1 Band structures (a) and the density of states (DOS) (b) of the SGS materials with parabolic-type dispersion. There are four different types of band structures or DOS.

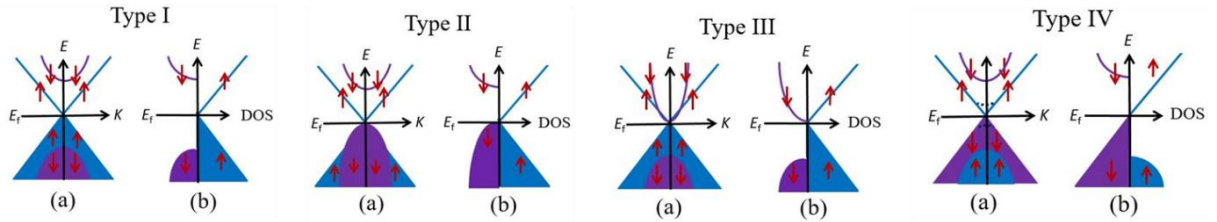

Figure 2 Band structures (a) and the density of states (DOS) (b) of the SGS materials with Dirac-type dispersion between energy and momentum. . There are four different types of bands or DOS.

1.2 SGSs Band structures and density of states

1.2.1 Parabolic dispersion

Type I: Zero gap for spin-up bands and a gap for spin down bands. Type II: Zero gap for spin up bands and a gap for spin down bands with top of spin-down valence band touches the Fermi level. Type III: Zero gap for spin-up and a gap for spin-down bands with bottom of spin-down conduction band touches the Fermi level. Type IV: Both spin-up and spin-down bands have a gap. But, the gap is zero for the spin-down and spin-up bands. The energy dispersion is parabolic for all the four types in both band structures and their DOS.

1.2.2 Linear dispersion or Dirac type SGSs.

Mathematically, except the linear dispersion, the arrangements for spin-up and spin-down bands should be the same as for parabolic SGSs. However, in real materials, some linear dispersion may not hold for a gapped state. Therefore, the energy dispersion with momentum or DOS are mixed in the Dirac type SGSs. For the type I: Zero gap for spin-up bands with linear dispersion and a gap for spin down bands with parabolic dispersion. Type II: Zero gap for spin-up and linear bands and a gap for spin-down parabolic bands with top of spin-down valence band touches the Fermi level. Type III: Zero gap for spin-up and a gap for spin-down bands with bottom of spin-down conduction band touches the Fermi level. Spin-down bands are parabolic. Type IV: Both spin-up and spin-down bands have a gap. But, the gap is zero for the spin-down and spin-up bands. All bands are parabolic. It is worth noting that the band structures of the SGSs could be either direct or indirect. Here we only discussed the case of direct bands.

Unique spin and charge states and novel effect in parabolic or Dirac SGSs. The SGSs significantly distinct from conventional semiconductors, insulators, metals. They also have different spin and charge features as compared to half-metals and recently discovered topological insulators, half Dirac semimetals, and Weyl metals. The main features of spin and charges are summarized as following: [1-3] Little energy is need for excitation; 2) Both fully spin polarized electrons and/or hole can be generated by light excitation or thermal activation; 3) both electrons and holes are fully polarized with same spin polarization (Type I) and they are massless for Dirac tape SGSs; 4) Conduction (type II) or valence (type III) bands are fully polarized, but valence band is spin degenerated or partially polarized; 5) Both conduction and valence bands are fully polarized, but they have opposite polarization; 6) Under the influence of perpendicular magnetic field, the spin fully polarized electron or holes can move to the edge of thin samples due to Hall effect.[2] 7). Emitted light from excitation should be polarized; 8) For symmetrical bands, a number of novel spin and charge related Hall effect are predicted.[3]

1.2 Quantum Anomalous Hall effect in SGSs

SGSs can be realized in a wide range of gapless or narrow band oxides and non-oxide semiconductors, ferromagnetic or antiferromagnetic semiconductors as indicated by their band structures. It was proposed in original proposal that fully spin polarized carriers can move to the edge of samples using Hall effect.[2] This feature is named as field-induced self-spin filtering effect.[2] If all spin polarized charge carriers move to the sample edge due to influence of either or external magnetic field, it turns to be a quantum Hall effect. A theoretical prediction of quantum anomalous Hall effect in an organic 2D ferromagnetic system which shows typical feature of type I Dirac SGSs.[5] This prediction on QAHE renders the SGSs for an alternative candidates for QAHE. Here are the basic principle: The bands structure for type I SGS shows that both conduction and valence bands are fully spin polarized with the same spin polarization. Speaking mathematically, for the case of direct bands, both top of valence band and bottom conduction bands touch exactly at the Fermi level. This is the case without considering additional interactions. Interestingly, when spin-orbital interaction is introduced, the spin fully polarized band will open a gap creating a topological edge state with a non-zero Chern number. So, the SGSs will turn into a Chern insulator with a dissipationless edge state—namely QAHE.[5] The realization of QAHE in both Dirac and parabolic SGSs are schematically shown in Figure 3.

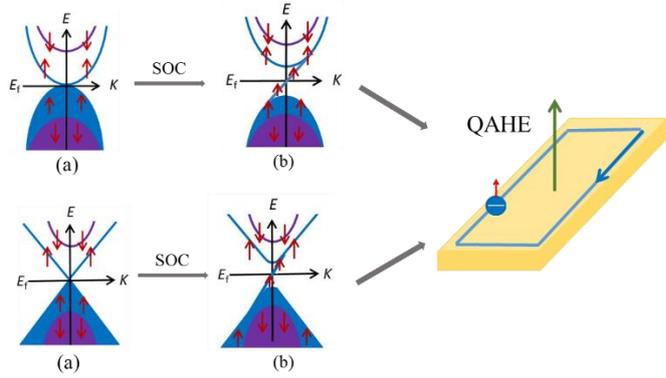

Figure 3. The band structures of parabolic and Dirac type SGS materials with spin orbital coupling, which leads to the quantum anomalous Hall effects.

Most of the predicted SGSs predicted or experimentally demonstrated so far. The Dirac type SGSs are particularly interesting. The Dirac-like linear SGS materials have massless carriers and low-dissipation transport properties. Figure 1 shows the electronic structures of the SGS materials with Dirac type dispersion. Quantum Anomalous Hall effects (QAHE) and new types of anomalous (quantum) Hall effect can be expected in these Dirac type SGS materials.[3] This type of SGS materials hold great potential for practical applications for ultrafast and low-energy consumption spintronic and electronic devices.[4]

## 2. SGS materials in Heusler compounds

Recently, fully-compensated SGS materials have been found, which has no net magnetization. Heusler compounds have high Curie temperature and are easy to fabricate. Some of them have already been confirmed as SGS materials. The SGS properties in this class of materials result from the Slater–Pauling curve. They include full-Heusler, half-Heusler, LiMgPdSn-type quaternary Heusler and $DO_3$-type compounds.[6]

2.1 Half-Heusler type SGS materials

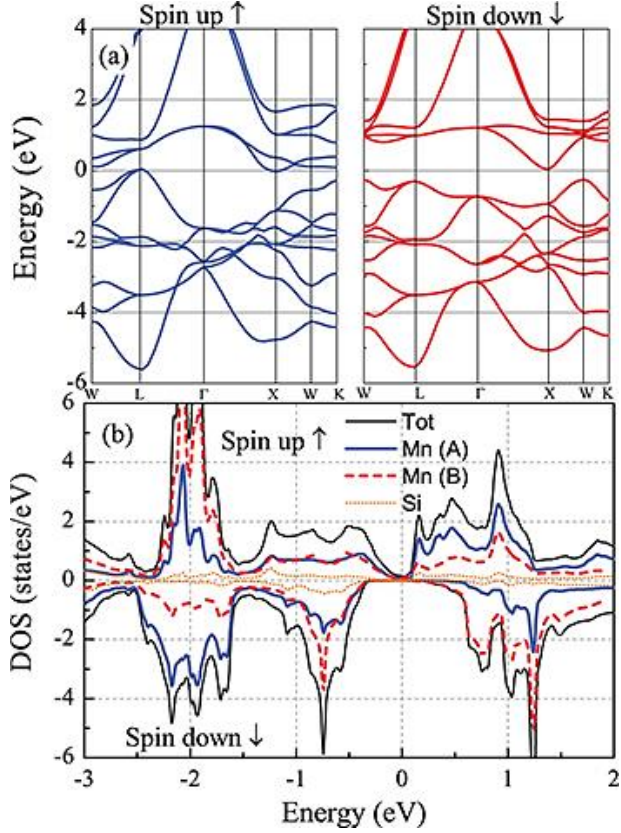

Figure 4 (a) The band structures and (b) the DOS in the Heusler compound $Mn_2Si$. Reproduced with permission [7]. Copyright (2015), European Physical Society.

The Heusler compounds $Mn_2Si$, $Cr_2ZnGe$, $Cr_2ZnSn$, $Ti_2VP$, and $Ti_2VSb_{0.5}As_{0.5}$ have been predicted as the candidates of SGS materials using first-principles calculations.[7] The electrons and holes in $Cr_2ZnGe$, $Cr_2ZnSn$, $Ti_2VP$, and $Ti_2VSb_{0.5}As_{0.5}$ are both fully spin polarized. In $Mn_2Si$, only holes are fully spin polarized, which can be considered as one member in half-Heusler family. The electronic-structure of $Mn_2Si$ is demonstrated in the Figure 4(a). The spin-up channel is zero-gap and the VB maximum touches the L-point of Fermi level. The CB minimum of spin-up and down electrons reach the X-point of Fermi level. There is no total magnetic moment in $Mn_2Si$ as the moments of Mn (A) are anti-parallel to Mn (B). The spin-up states are nearly fully occupied and the spin-down states are partially unoccupied, which leads to a net moment of spin-up electrons in the DOS of Mn(B).[7]

In additions, the electronic structure and magnetic properties have also been studied in $Mn_2Si$ by element doping.[8] The study shows that $Mn_2Si_{1-x}Ge_x$ compounds are compensated ferrimagnets and half-metallic. The band structure, Fermi level and magnetic moment could be changed by doping while the half-metallic and ferrimagnetic property are not destroyed. Furthermore, the SGS features may disappear when Si is replaced by Ge in the $Mn_2Si$ compound.

2.2 Full-Heusler type SGS materials

2.2.1 $Mn_2CoAl$ compound

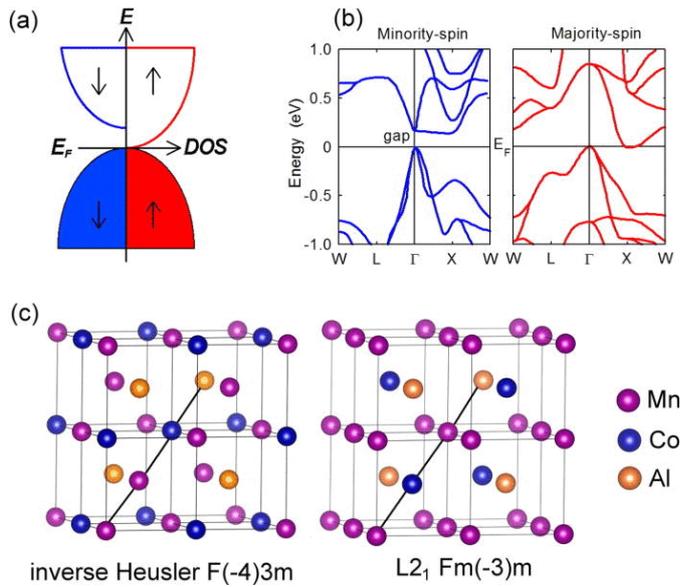

Figure 5 (a) The DOS of a SGS material. (b) Band structures of the Mn$_2$CoAl compound. (c) Unit cell structures of inverse-Heusler and the L2$_1$ structures. Reproduced with permission [9]. Copyright (2013), AIP Publishing.

Some full-Heusler compounds, such as Mn$_2$CoAl, have been identified as SGS materials. (already mentioned this in page 5)[9] The magnetic, transport and structural properties of Heusler ferrimagnet Mn$_2$CoAl films have been investigated.[9, 10] The anomalous Hall resistivity of Mn$_2$CoAl compound is in proportion to the longitudinal resistivity's square. They have the magnetism that results from the topological Berry curvature. Figure 5(a) shows that the spin-down states have a gap while the spin-up states have no gap. Figure 5(b) demonstrates that the VB maximum of spin-up bands touch at Γ-point of Fermi level and the CB minimum touch at X-point of Fermi level. The VB maximum of the spin-down bands touch the Γ-point of Fermi level while the CB minimum located at 0.2 eV above the X-point of Fermi level.

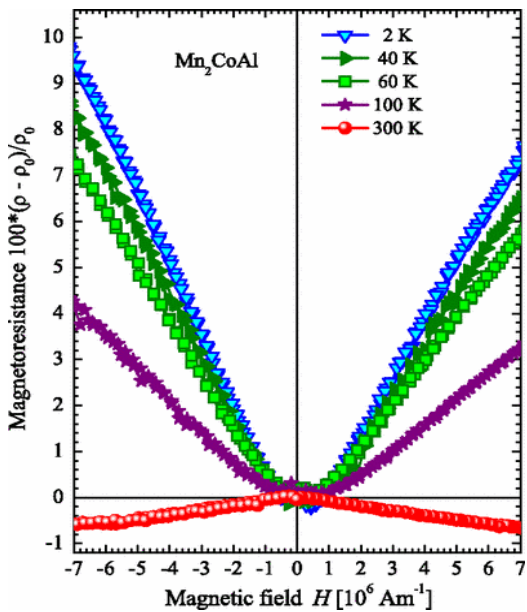

Figure 6. Magnetoresistance of Mn$_2$CoAl measured at different temperatures. Reproduced with permission [11]. Copyright (2013), American Physical Society.

The magnetic and electronic properties of Mn$_2$CoAl compound were investigated using first-principles calculations.[12] The calculated band structures of Mn$_2$CoAl have the SGS features. The exchange constant calculations display that the magnetic states are stable and the interactions in Mn$_2$CoAl are short-range owing to the strong interactions of nearest neighbours. The calculated Curie temperature in Mn$_2$CoAl compound is 890 K.[13] The magnetoresistance (MR) of Mn$_2$CoAl compound has been measured at different temperatures, as shown in Figure 6. The MR is found to be linear in low magnetic field region.[11]

Mn$_2$CoAl thin films have been fabricated on the GaAs substrate by the molecular beam epitaxy (MBE) and they exhibited a metallic-like electronic property at low temperatures.[9] The calculated Curie temperature is 550 K in epitaxial Mn$_2$CoAl films. The magnetotransport properties of the Mn$_2$CoAl films were studied and compared with the bulk of the Mn$_2$CoAl compound. In contrast to the Mn$_2$CoAl films on GaAs, this film revealed semiconductor-like behaviour over the whole temperature range.[10] In addition, electronic, magnetic, elastic, structural, thermodynamic, and disorder properties of the Mn$_2$CoAl compound have also been studied by the first-principles calculations and experimental methods.[14]

2.2.2 Ti$_2$CoSi and Ti$_2$MnAl compounds

Based on the first-principles calculation method, Ti$_2$CoSi and Ti$_2$MnAl compounds were predicted to be SGS materials.[15] The calculated total magnetic moment for Ti$_2$MnAl and Ti$_2$CoSi compounds are 0 $\mu_B$ and 3 $\mu_B$ respectively. For the electronic structure, Ti$_2$MnAl is gapless at the midpoint of the L and W points. Ti$_2$CoSi has indirect zero-gap. The CB minimum of spin-up states locates at the X-point while the VB maximum of spin-up states locates at the Γ point. Ti$_2$MnAl compound combines the features of SGS materials including zero net spin moment and quite high Curie temperature.

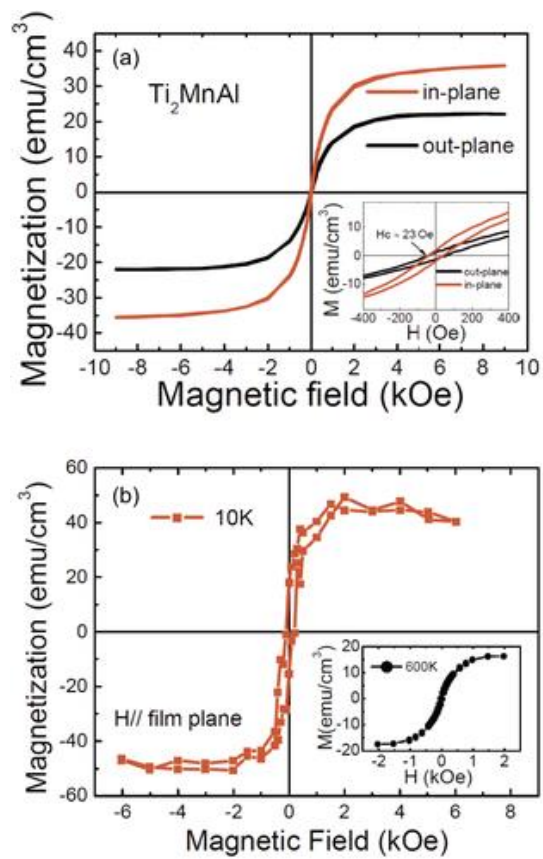

Figure 7. (a) The magnetization for the Ti$_2$MnAl/Si(001) film as a function of the magnetic fields at room temperature. The inset shows the details of the low field range. (b) Magnetic fields dependent magnetization at the temperatures of 10 K and 600 K. Reproduced with permission [16]. Copyright (2015), John Wiley and Sons.

The Ti$_2$MnAl thin films have been deposited on the Si (001) substrates using magnetron sputtering technology.[16] The magnetic and electronic transport properties were investigated and the material was found to exhibit semiconducting behaviour above 70 K. The films have the metallic behaviour between 70 K and 15 K. The change in the resistivity may suggest that the Ti$_2$MnAl compound is a SGS material. Magnetic fields and temperatures dependent magnetism has also been measured, as shown in Figure 7. The resistance and magnetization show a linear temperature dependence above 100 K. All this behaviour demonstrates that the Ti$_2$MnAl film is a SGS material.

2.2.3 Ti$_2$CrSi and Ti$_2$CrSn compounds

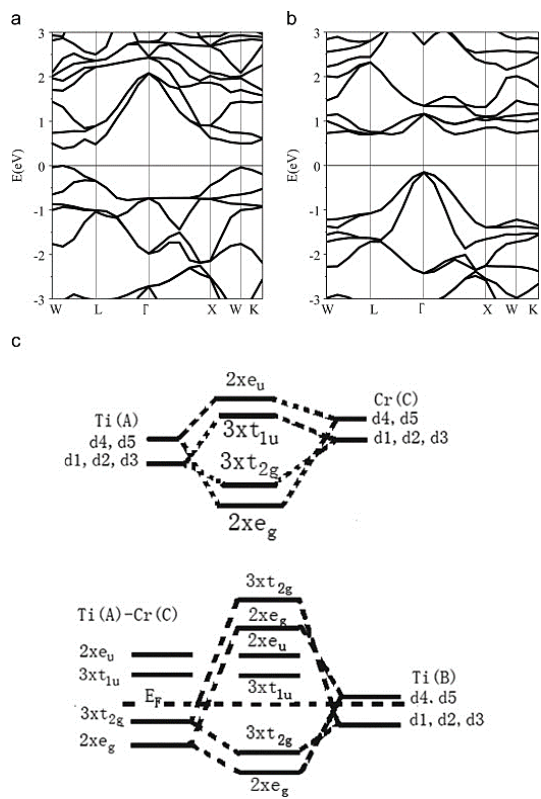

Figure 8. Band structures of (a) the spin-up and (b) the spin-down channels of the Ti$_2$CrSn compound. (c) Views of hybridizations of the spin-down channels. Reproduced with permission [17]. Copyright (2014), Elsevier.

Ti$_2$CrSi and Ti$_2$CrSn compounds have also been studied as SGS materials.[17, 18] The band structure is shown in Figure 8. In the Ti$_2$CrSi compound, tetragonal distortion and uniform strain have been used to modulate the band structures.[19] The Ti$_2$CrSi compound shows the SGS features at a tetragonal distortion. Doping effects can be used to realize SGS properties through modulating the band gaps of two spin channels. When Sn is partly replaced by Si and Ge, the band gap value of the majority spin channel decreases. The Ti$_2$CrSn$_{0.5}$Si$_{0.5}$ is a SGS materials and the band gap is zero in the majority spin channel while it is 0.57 eV in the minority spin channel. In the Ti$_2$CrSn$_{0.5}$Ge$_{0.5}$ compound, the values of the band gaps in the minority spin channel and the majority spin channel are 0.56 eV and 0.07 eV, respectively. Hence, it can be classified as a SGS material.

2.3 Binary DO$_3$-type Heusler Cr$_3$Al and V$_3$Al compounds

When the lattice constant is enlarged, the Cr$_3$Al compound could transit into a SGS material from a metallic material.[20] The equilibrium constant is 5.92 Å for both the minority and majority spin channels intersect with the Fermi level. The DOS for the minority spin channel transits to a lower energy region while the DOS for the majority-spin channel reaches the Fermi level when the lattice constant is 0.6 nm. The calculated total magnetic moment is about 3.0 μ$_B$ per formula unit and the Cr$_3$Al compound becomes a SGS material.

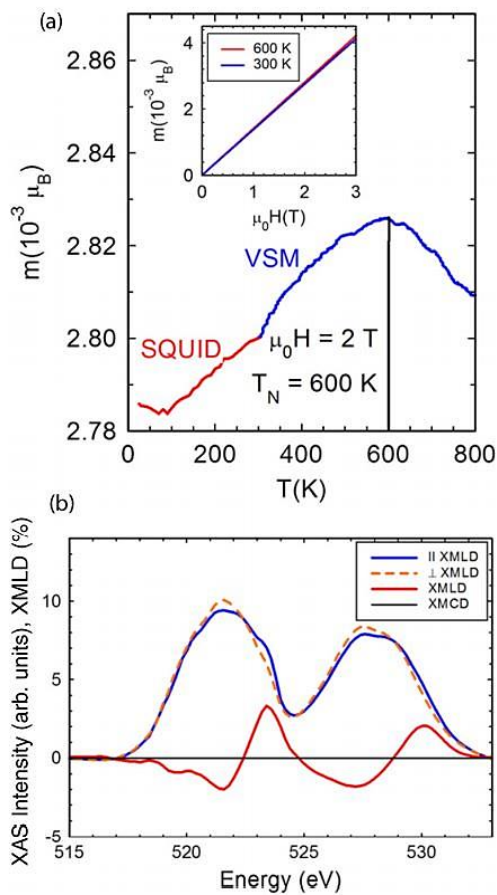

Figure 9 (a) The magnetic moment of V$_3$Al. The inset figure demonstrates the moment as a function of applied field at 300 and 600 K. (b) X-ray absorption spectroscopy (XAS) of the V atoms in the V$_3$Al compound. The X-ray linear magnetic dichroism (XMLD) intensity displays 5% difference of two XAS spectra. In contrast, the X-ray magnetic circular dichroism (XMCD) polarization at zero intensity shows a small magnetic moment on the V atoms. Reproduced with permission [21]. Copyright (2015), American Physical Society.

The V$_3$Al compound was also predicted to be a SGS materials based on the first principle calculations.[15] Both the minority and majority spin bands has no bang gap. The V$_3$Al compound is a gapless semiconductor with anti-ferromagnetism, as shown in Figure 9. It was fabricated using arc-melting and annealing methods.[21] The synchrotron measurements was used to confirm the antiferromagnetic behaviour, and the magnetic anomaly was observed on different V atoms.

2.4 Quaternary Heusler CoFeCrGa and CoFeMnSi compounds

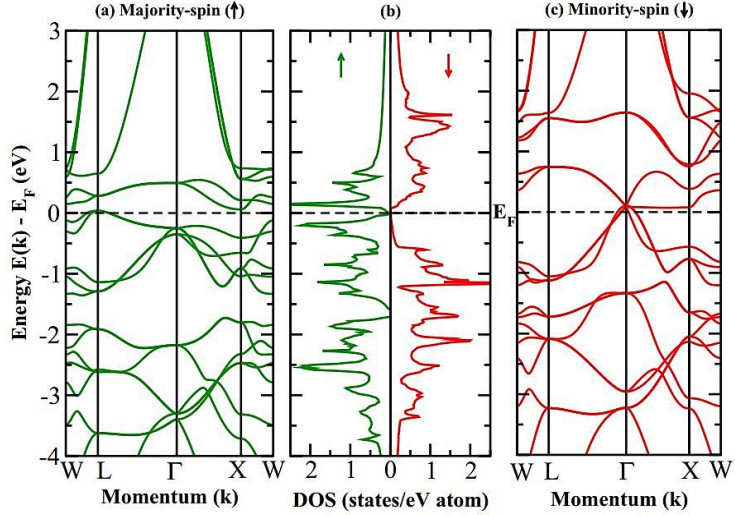

Figure 10. Band structures of (a) the majority spin states, (b) DOS, and (c) the minority spin states of the CoFeCrGa compound. Reproduced with permission [22]. Copyright (2015), American Physical Society.

The quaternary Heusler CoFeCrGa compound has been proposed as a SGS material from first-principles calculations and experiments.[22] The DOS and band structures are shown in Figure 10. The zero-gap feature of the majority spin state and non-zero gap feature of the minority spin state suggest that this compound owns the SGS property. This material has been investigated by Mössbauer spectroscopy, magnetization and electronic transport experiments. This compound can change from a SGS material to a half-metal under pressure due to the unique electronic structure. The saturation magnetism was found at 8 K and the Curie temperature is up to 400 K. The carrier concentration and electrical conductivity are nearly independent to temperatures. At 5 K, the estimated anomalous Hall coefficient is about 185 S/cm.

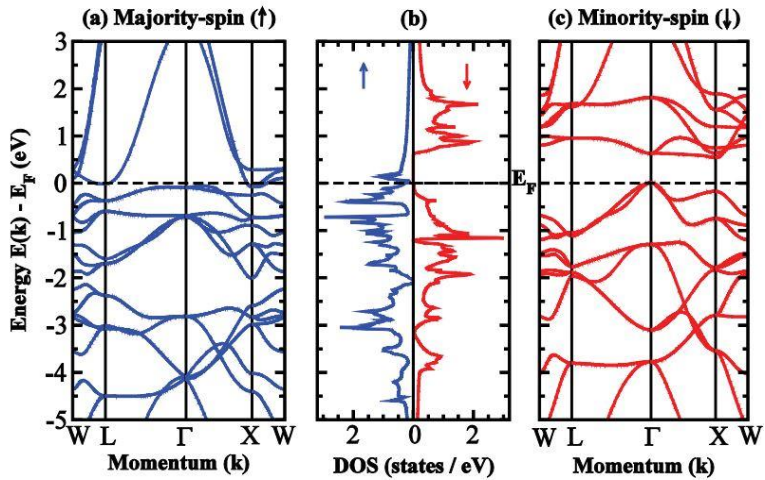

Figure 11 Band structure and DOS in the CoFeMnSi compound: (a) the majority spin bands, (b) the DOS, (c) the minority-spin bands. Reproduced with permission [23]. Copyright (2015), American Physical Society.

The band structure of the Heusler CoFeMnSi compound has been investigated by the first-principle calculations.[23, 24] The study indicates that the compound owns the SGS eature. Figure 11 shows the band structure with spin-polarization and the DOS. The DOS has a band gap of ~0.62 eV in one spin sub band while another spin subband has a small ban gap. The SGS feature of this compound was studied based on the experimental measurements of its magnetization, spin polarization, structural, and transport properties.[23] The compound has a cubic Heusler structure and the saturation magnetization is about 3.7 μB/f.u. The Curie temperature is approximately 620 K. The saturation magnetization follows the Slater-Pauling rule. From 5 K to 300 K, the electrical conductivity and carrier concentration are nearly independent to temperatures. At 5 K, the anomalous Hall coefficient is 162 S/cm. The spin-polarization value is about 0.64 by Point contact Andreev reflection measurements. These results strongly support that the CoFeMnSi compound is a SGS material. The structural, electronic, and magnetic properties of CoFeMnSi thin films on GaAs(001) have also been investigated.[25] The heterostructure with a MnMn-terminated interface holds fully spin polarization.

## 3. Dirac type SGS materials

Based on the band structures, SGS materials can also be divided into Dirac-like linear and parabolic SGSs. The Dirac SGS materials have massless carriers and low-dissipation electronic transport. Members of this class of SGSs are regarded as promising material candidates for low-energy loss spintronic and electronic devices.[4] [26] Dirac type SGS materials have two types: $p$-state type and $d$-state type, based on the origin of the Dirac states. Dirac SGSs have many characteristics including fully spin-polarization, massless carriers, high Fermi velocity, and low-transport dissipation.

3.1 $d$-state Dirac SGS materials

3.1.1 Mn-intercalated graphene on SiC substrate

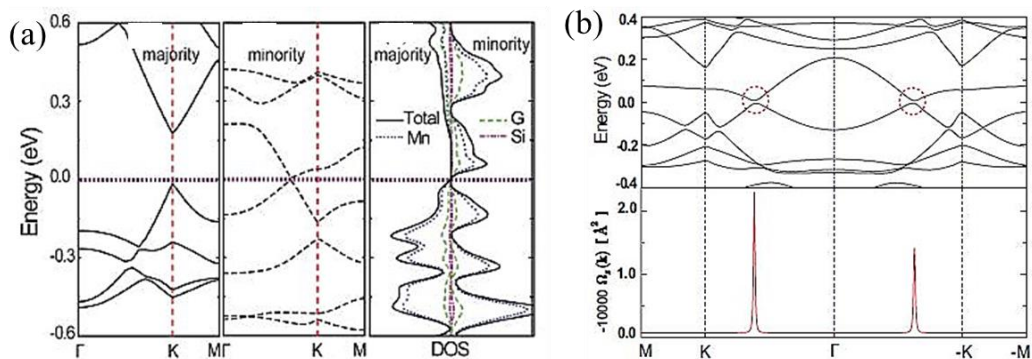

Figure 12 (a) Band structure and the DOS of the Mn-intercalated graphene with coverage, χ = 1/3 monolayer (ML) Mn. (b) Band structure with SOC and Berry curvature along high symmetry line of Mn-intercalated graphene on SiC. Reproduced with permission [27]. Copyright (2015), American Physical Society.

Mn-intercalated graphene was predicted to be a SGS material based on the first-principles calculations.[27] In the structure, Mn atoms are between the graphene and the SiC (0001) substrate. The Dirac type SGS state only appears in a fixed Mn coverage region and can be tuned by substrate. These are relevant to the Mn-SiC interaction and 2D-like symmetry of inversion. As shown in Figure 12, the gap of majority-spin channel is 0.19 eV, the minority-spin channel has a Dirac cone and gapless property. The DOS shows the main contribution of Dirac states results from the Mn atoms and, hence, it is a $d$-state Dirac SGS material. The SOC can open a gap in the Dirac cone of minority-spin channel and lead to the QAH edge state. By integration over the Brillouin zone, an odd Chern number C = 1 can be obtained, which confirm that one chiral state can exist on the edge in the ribbon structure.

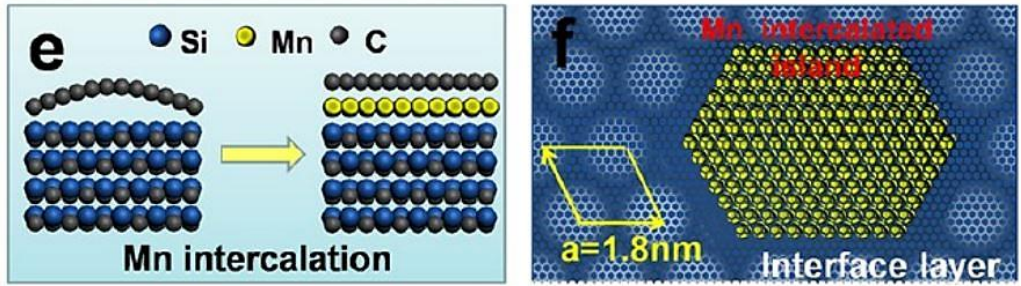

Figure 13 Schematic draw of Mn intercalation in SiC and the top view of the interface layer. Reproduced with permission [28]. Copyright (2012), American Chemical Society.

Figure 13 shows the view of the epitaxial graphene/Mn/SiC(0001) structure which is formed through intercalating Mn atoms into graphene surfaces.[28] The Dirac states in the structure were confirmed through Mn coverage tuning. The Dirac point lowers when the graphene samples have Mn intercalation. The Dirac point moves down with Mn coverage tuning and vanishes when the coverage $\chi = 0.6$ monolayer.

### 3.1.2 $CrO_2$/$TiO_2$ heterostructures

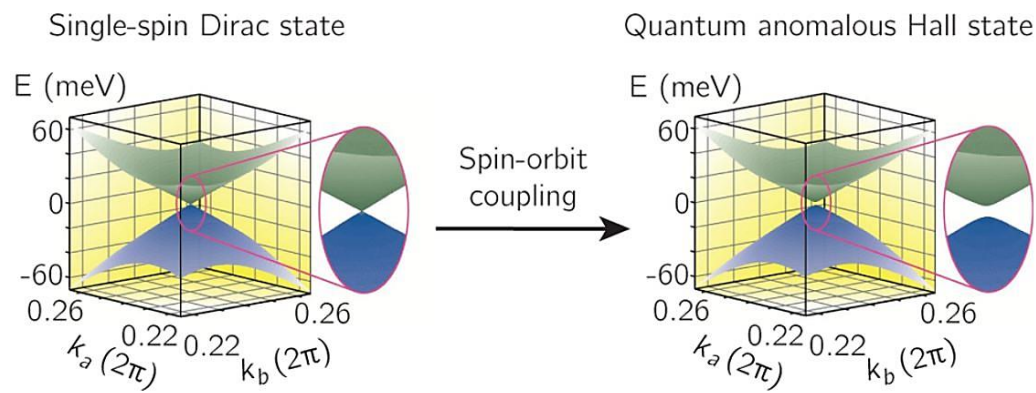

Figure 14. Dirac point in the $(CrO_2)_4$/$(TiO_2)_{10}$ superlattice. (a) The band structure without the spin orbital coupling (SOC). (b) The band structure with SOC with $c$-oriented magnetization. Reproduced with permission [29]. Copyright (2015), American Chemical Society.

Superlattices and heterostructures of the $CrO_2$ the $TiO_2$ have been found to be Dirac type SGS systems.[29] The single-spin Dirac states can be generated from the interfacial electrons at the $CrO_2$/$TiO_2$ heterostructures. The Dirac state results from the 2D $CrO_2$ layer and is determined by the $d$-orbitals of Cr atoms. Hence, the system belongs to the $d$-state type. When the SOC is included, a QAH phase with Chern number of $\pm 2$ can be obtained in the superlattice, as shown in Figure 14. Therefore, they can be useful for designing dissipation-less electronic and spintronic devices. Besides, when symmetry-allowed SOC terms were included, the Hamiltonian gives a Chern number of $\pm 2$ and two gapless chiral edge within the bulk gap. This will lead to the QAHE in this system.

### 3.1.3 Metal-organic framework SGS materials

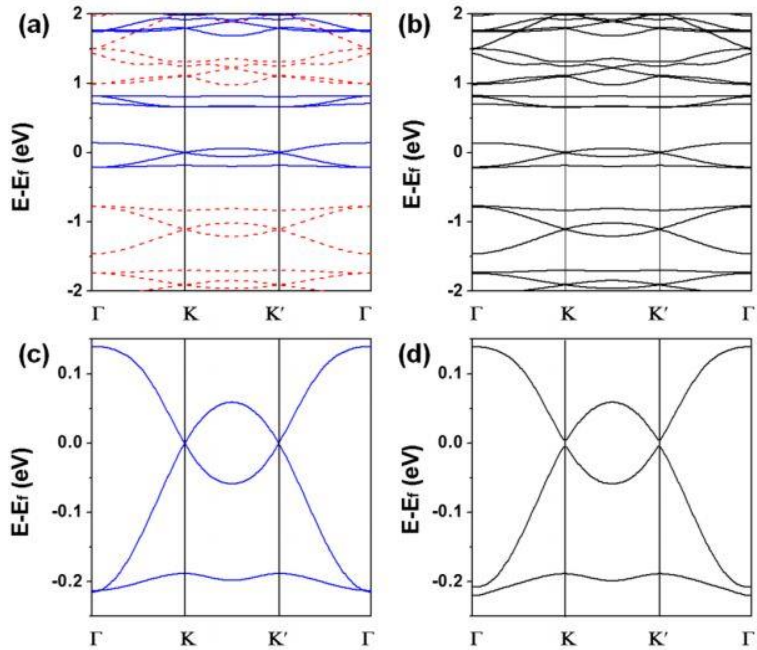

Figure 15. The band structures of the Mn($C_6H_5$)$_3$ compound. Band structures of the Mn($C_6H_5$)$_3$ lattice without and with SOC, respectively. Red dashed lines and blue solid lines denote spin-up and spin-down bands. (c), (d) Magnification of (a) and (b) around the Fermi level, respectively. Reproduced with permission [5]. Copyright (2013), American Physical Society.

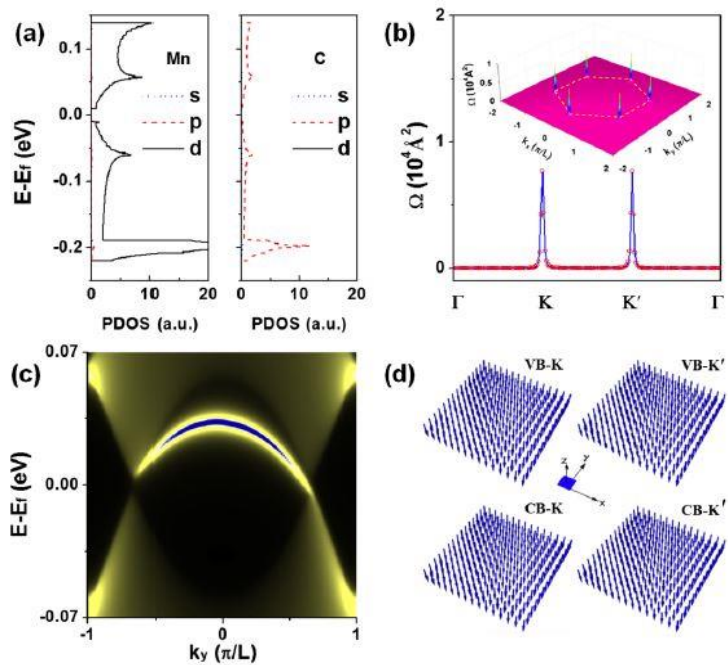

Figure 16 (a) Density of states (DOS) of Dirac bands around the Fermi level. (b) 2D distribution of berry curvature for the valence band in the momentum space. (c) Edge states inside the Dirac gap of the TMn lattice. (d) 3D spin texture for the highest occupied valence band and the lowest unoccupied conduction band. Reproduced with permission [5]. Copyright (2013), American Physical Society.

(1) Mn($C_6H_5$)$_3$ compound

Several metal-organic framework have been predicted as the SGS materials. One of them is the Mn($C_6H_5$)$_3$ compound, which has a hexagonal lattice. As shown in Figure 15, two Dirac cones are in its spin-down channel and it has gapless features.[5] The two Dirac cones have high symmetry and the Dirac points are located at the Fermi level. There are also two Dirac cones below the Fermi level in the spin-up channel. The Dirac states mostly originate from the *d*-orbitals of Mn and, hence, it is a *d*-state Dirac SGS. The nonzero Chern number can lead to the chiral edge states within Dirac gap. The local edge states in Mn($C_6H_5$)$_3$ were calculated and shown in Figure 16c. The number of the edge states shows the value of the Chern number. The edge state will lead to the realization of QAHE in this system.

(2) $Ni_2C_{18}H_{12}$ and $Co_2C_{18}H_{12}$ compounds

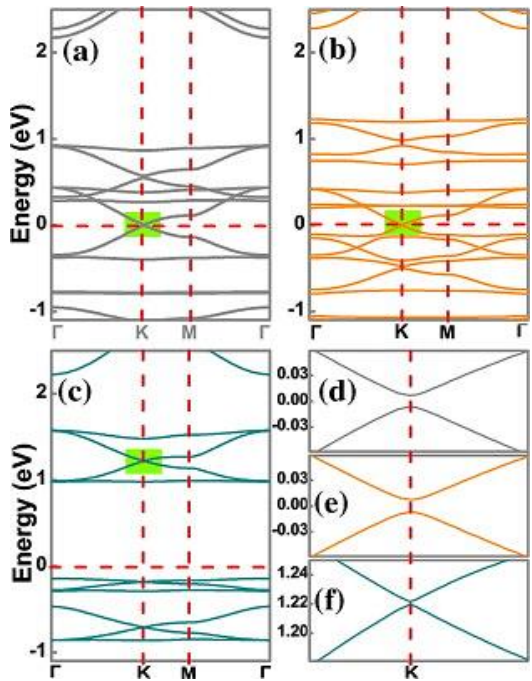

Figure 17. Band structures of (a) $Ni_2C_{18}H_{12}$ compound, (b) $Co_2C_{18}H_{12}$ compound, and (c) high-$Co_2C_{18}H_{12}$ compound with SOC. (d–f) Views of band structure around the Dirac points. Reproduced with permission [30]. Copyright (2014), Elsevier.

The Density functional theory (DFT) shows that the 2D $Co_2C_{18}H_{12}$ compound is spin-polarized.[30] The 2D $Ni_2C_{18}H_{12}$ and the 2D $Co_2C_{18}H_{12}$ have half-metallic Dirac points close to the Fermi level. One spin channel has Dirac points at the Fermi level, as displayed in the Figure 17. Based on DOS analysis, the Dirac states are from *d*-orbitals so that they are *d*-state Dirac SGS materials. Besides, they both have robust ferromagnetism. When the SOC is added, a band gap may be opened in the Dirac cone. Therefore, they hold the potential for the realization of QAHE.

(3) $Ni_2C_{24}S_6H_{12}$ compound

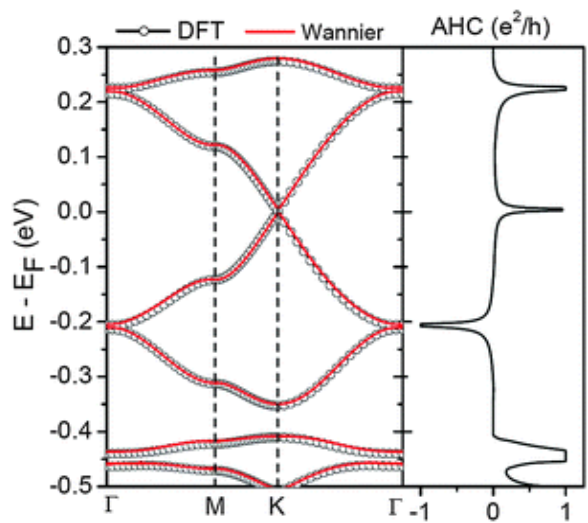

Figure 18. (Left) The electronic band structures close to the Fermi level from the DFT calculations and Wannier functions. (Right) The calculated anomalous Hall conductivity. Reproduced with permission [31]. Copyright (2016), Royal Society of Chemistry.

Dirac cones with spin-polarization and non-trivial topological property have been observed in the metal–organic framework $Ni_2C_{24}S_6H_{12}$, as shown in Figure 18.[31] The topological non-trivial states of the homologous kagome lattices result from its kagome bands. They connect the profiles of the four bands in a ruby lattice. $Ni_2C_{24}S_6H_{12}$ has ferromagnetism and the Curie temperature is up to 630 K, calculated from the Monte Carlo simulation. When the temperatures increase from 600–630 K, the magnetic moment of per unit cell decreases from 1.6 $\mu_B$ to 0 $\mu_B$. When the SOC is added, band gaps can be created in at Dirac point and $\Gamma$ point in Brillouin zone. The nontrivial topology of the SOC induced band gaps can be indentified by Chern numbers. The calculated Chern numbers in this compound is C = 1 and -1. This makes $Ni_2C_{24}S_6H_{12}$ a Chern topological insulator and a candidate for realizing QAHE.

(4) $Mn_2C_6S_{12}$ compound

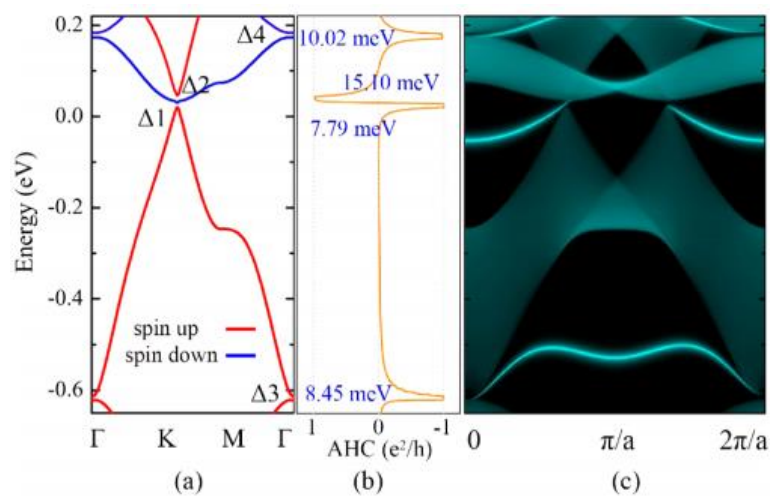

Figure 19 (a) Band structure for $Mn_2C_6S_{12}$ compound with SOC. (b) Calculated anomalous Hall coefficient and (c) semi-infinite edge states. The Fermi level energy is zero. Reproduced with permission [32]. Copyright (2017), American Chemical Society.

Chern half-metal and Chern insulator states have been observed in 2D SGS material $Mn_2C_6S_{12}$ compound.[32] First-principles calculations showed the Dirac point in the channel of spin-up crosses the Fermi level and the gap in the channel of spin-down is about 1.53 eV. The Dirac state is the *d*-state type as it is the contributions of *d*-orbitals of Mn atoms. Specially, the parabolic dispersion and spin-polarized Dirac cone coexist in the $Mn_2C_6S_{12}$ compound. In addition, the $Mn_2C_6S_{12}$ holds stable ferromagnetism at room temperature from the Monte Carlo simulations. When the SOC is considered, a band gap is opened in Dirac cone and the degenerate points move up, which leads to multiple topologically non-trivial states, as shown in Figure 19. The values of type-I and type-II band gaps are between 7.79 meV and 15.10 meV due to Due to the inter-spin and intra-spin SOC. The nanoribbons edge states and non-zero Chern numbers confirmed the topological non-trivial phase of these band gaps, which suggest that $Mn_2C_6S_{12}$ may be a Chern semi-metal and insulator. Figure 19c shows the calculated topologically nontrivial edge states between the bulk states, which is the feature of Chern insulators. The bulk SOC band gaps are separated with edge states and will led to the achievements of the QAHE at high temperature. The unique electron properties provide the opportunities for the design of low-energy electronic devices.

3.2 Transition metal halides

3.2.1 $VCl_3$ and $VI_3$ monolayers

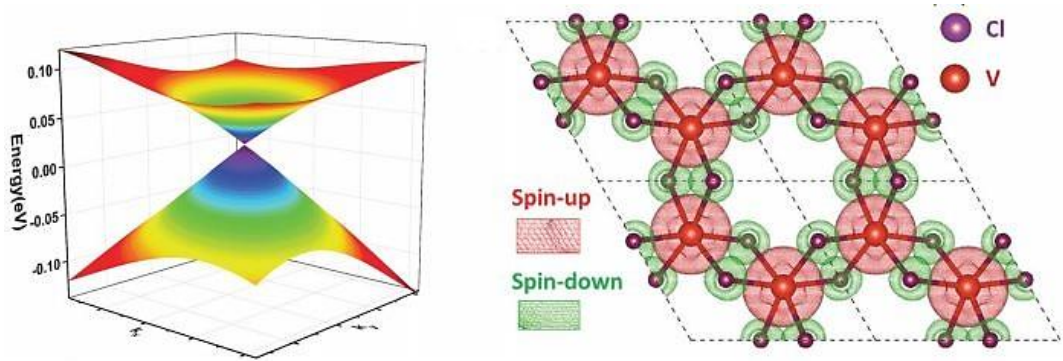

Figure 20. (Left) The 3D Dirac band structure of $VCl_3$ and (right) the spin-polarized charge density of $VCl_3$. Reproduced with permission [33]. Copyright (2016), Royal Society of Chemistry.

The DFT approach was used to study the magnetic and electronic structures as well as stability of the $VCl_3$ and $VI_3$ monolayers.[33, 34] [35] The small cleavage energies are the basis for the exfoliation from their layered bulk. The phonon calculations show that $VI_3$ and $VCl_3$ monolayers and they can act as free-standing 2D materials. Band structure calculations show that they hold their energy dispersions is linear at Dirac points above Fermi level.[33] Their electrons have a gap in one spin channel and high mobility in another spin channel. The corresponding bands around the Fermi level is 3D. The spin-polarized charge densities of the $VCl_3$ monolayers are presented in Figure 20. They also have half-metallicity and intrinsic ferromagnetism. The Curie temperatures of $VI_3$ and $VCl_3$ sheets are only 98 K and 80 K calculated from Monte Carlo simulations. But, by carrier doping, their Curie temperature could be as high as room temperature. The Dirac points of $VI_3$ and $VCl_3$ monolayers result from the V*d* electrons. They have large SOC related gaps of 12 meV and 29 meV for $VI_3$ and $VCl_3$ monolayers, respectively. This could lead to the realization of high temperature QAHE in this system.

3.2.2 $NiCl_3$ monolayer and $PtCl_3$ nanosheet

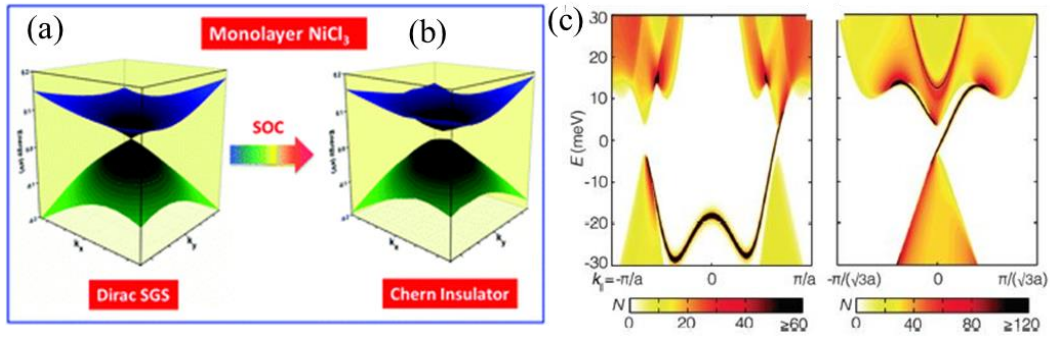

Figure 21 (a-b) Dirac SGS and Chern insulator (with SOC): NiCl$_3$ monolayer. (c) Calculated local density of states of edge states for zigzag and armchair insulators. Reproduced with permission [36]. Copyright (2017), Royal Society of Chemistry.

The NiCl$_3$ monolayer was found to have SGS features and ferromagnetism with $T_c$ of up to 400 K.[36] As shown in Figure 21, the NiCl$_3$ monolayer can transfer to a Chern insulator that has a gap of ~24 meV after adding SOC interaction. The structure of the NiCl$_3$ monolayer has robust thermally stable up to the room temperature. The magnetism in the NiCl$_3$ is also robust at room temperature and the magnetic moment of per unit cell is about 2 $\mu_B$. The $d$-state type SGS with SOC has larger gaps than the $p$-state type with SOC. The edge states can be created between the bulk band gaps. The warmer colors (darker) represent higher local density of states at the edge. Therefore, the NiCl$_3$ monolayer with a $d$-state Dirac SGS is a good platform to achieve QAHE at high temperature. The QAH state can result in a chiral edge states, which indicate the topological non-triviality of NiCl$_3$ monolayer. Hence, the NiCl$_3$ monolayer has potential for the QAHE and practical applications in the dissipationless spintronics.

3.2.3 MnF$_3$, MnCl$_3$, MnBr$_3$, and MnI$_3$

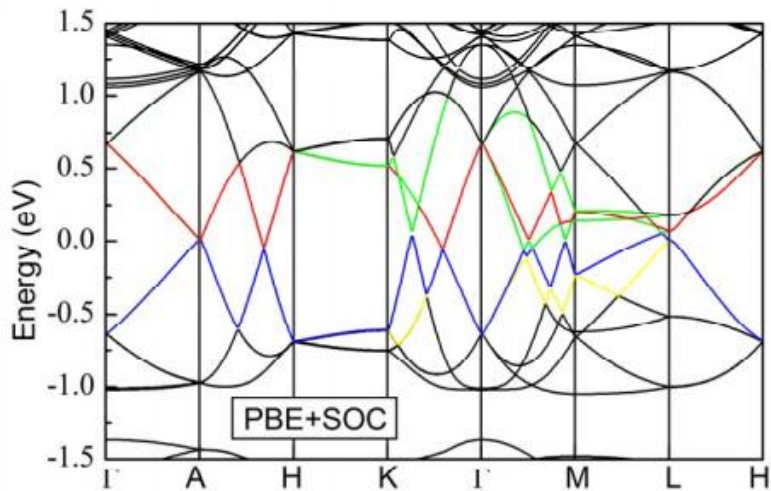

Figure 22. Band structure of MnF$_3$ obtained from the Perdew-Burke-Ernzerhof (PBE) method with SOC. The Fermi level is zero. Reproduced with permission [37]. Copyright (2017), American Physical Society.

Manganese trihalides, including MnF$_3$, MnCl$_3$, MnBr$_3$, and MnI$_3$, have been predicted to be two-dimensional SGS materials.[37, 38] The MnF$_3$ has a big gap in one spin channel and has multiple Dirac cones in another spin channel, as shown in the Figure 22.[37] The calculated Fermi velocity of the Dirac cones is similar with graphene. The band structure also shows that the MnF$_3$ has rings of Dirac nodes. In additions, the MnF$_3$ also has lots of novel properties, such as multiple Dirac rings, high carrier mobility and large spin polarization. The MnF$_3$ has great potential in ultrafast spintronics because the speed of the electrons and holes with spin-polarization is much faster than that in other SGS materials, such as PbPdO$_2$, Mn$_2$CoAl and diluted magnetic semiconductors.

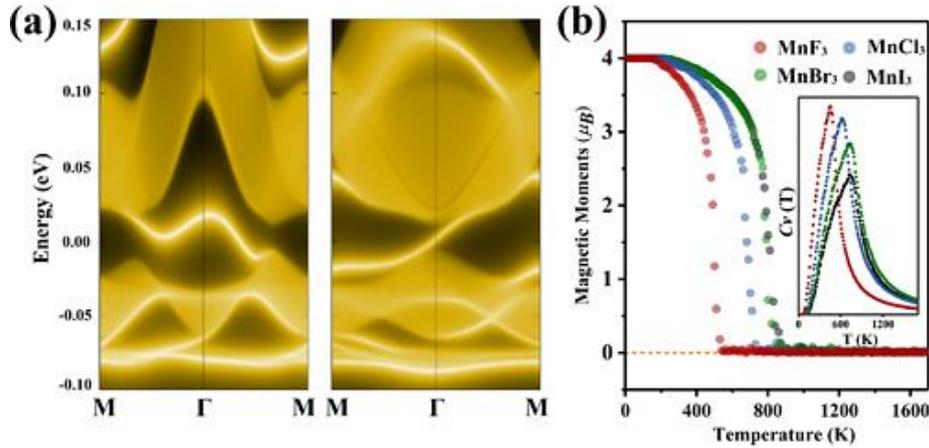

Figure 23 (a) Band structure of a MnBr$_3$ ribbon with zigzag (left) and armchair (right) edges. (b) Temperature dependent Mn$^{3+}$ magnetic moment with temperature dependent heat capacity (inset) of 2D MnF$_3$, MnCl$_3$, MnBr$_3$ and MnI$_3$ trihalides. Reproduced with permission [38]. Copyright (2018), American Physical Society.

2D MnCl$_3$, MnBr$_3$, and MnI$_3$ have also been studied, and they all have a Dirac cone in one spin channel with very high mobility and a large gap in another spin channel.[38] They are stable at high temperatures and hold large magnetic anisotropy energy, large magnetic moments, and high Curie temperatures, as shown in Figure 23. They can be obtained by simple exfoliation, similar to other van der Waals crystals. Besides, the SOC can induce a band gap and possible QAH state.

3.3 *p*-state Dirac type SGS materials

3.3.1 g-C$_{14}$N$_{12}$ and g-C$_{14}$N$_{12}$ compounds

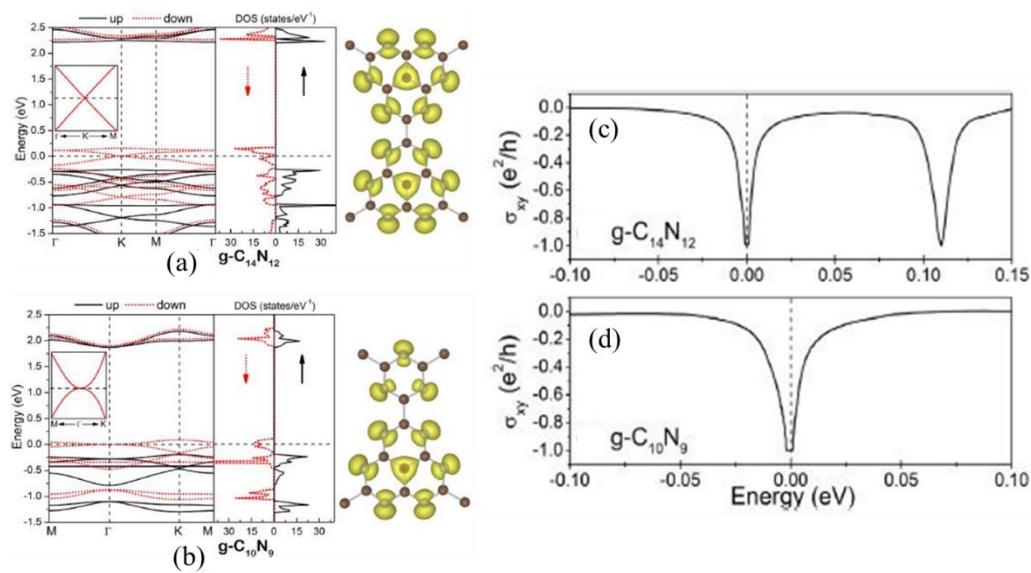

Figure 24 Band structure, DOS and the electron density distribution for (a) the g-$C_{14}N_{12}$ and (b) the g-$C_{10}N_9$. (c) The anomalous Hall conductivity of g-$C_{14}N_{12}$ (c) and g-$C_{10}N_9$ (d) with the Fermi energy set to zero. Reproduced with permission [39]. Copyright (2015), Elsevier.

Graphitic carbon nitrides have been predicted as *p*-state type Dirac SGS materials from first principles.[39] The electrons or holes in the g-$C_{14}N_{12}$ behave as spin-polarized massless Dirac fermions and the band structure is linear dispersion (Figure 24a). The electrons and holes in the g-$C_{10}N_9$ have different effective masses and the band structure is parabolic dispersion (Figure 24a). The magnetic moment for g-$C_{14}N_{12}$ is 1.0 $\mu_B$ per unit cell. The magnetic moment for g-$C_{14}N_{12}$ is 2.0 $\mu_B$ per unit cell. These magnetic moments have ferromagnetic configurations and form a stable ferromagnetic state. The Curie temperature is up to 830 K for g-$C_{14}N_{12}$ and is 205 K for g-$C_{10}N_9$, respectively. Moreover, based on the calculation, the SOC gap can be opened between the two Dirac bands at the Fermi level. The Chern number of the two SOC gaps has an integer value of -1, which means SOC gaps are topologically nontrivial. The Figure 24 shows the anomalous Hall conductivity in the two compounds. These confirm that the systems are promising platforms to realize the QAHE state in such as non-metal materials.

3.3.2 $YN_2$ and $MoN_2$ monolayers

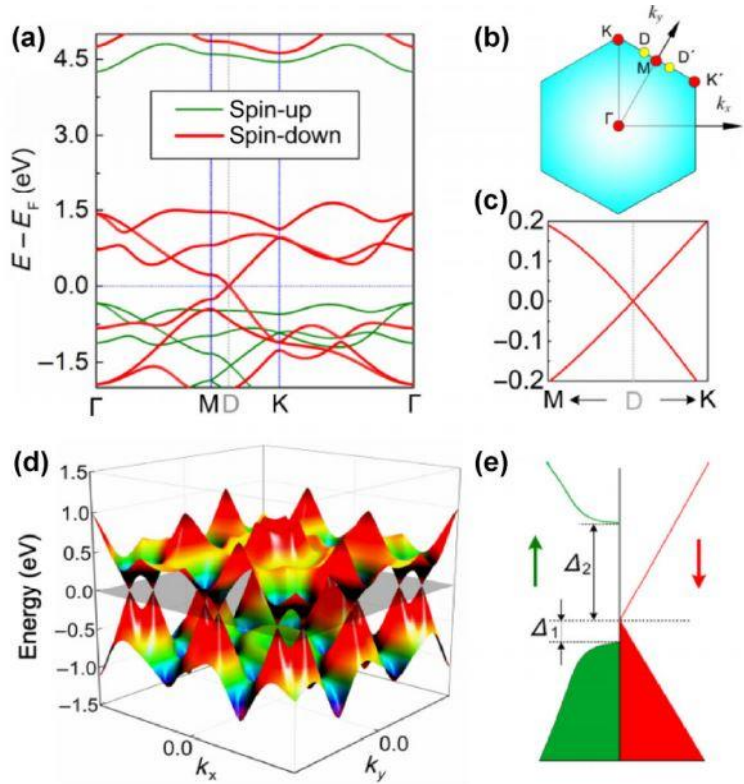

Figure 25 (a) Band structure in the 1T-$YN_2$ monolayer. (b) The 2D Brillouin zone. (c) Linear dispersion of the band structure. (d) 3D view of the spin-down channel. (e) B and structure of 1T-$YN_2$ monolayer. Reproduced with permission [40]. Copyright (2016), Springer Nature.

$YN_2$ monolayer has been predicted to be a *p*-state SGS material that would be suitable for high-speed spintronics.[40] The DFT calculations shows that the $YN_2$ monolayer has thermal and mechanical stability. The $YN_2$ monolayer holds spin-polarized Dirac states and very high Fermi velocity, as shown in Figure 25. It is ferromagnetic and the Curie temperature is ~332 K. What's more, it is difficult to destroy the ferromagnetic ground state by carrier doping and external strain.

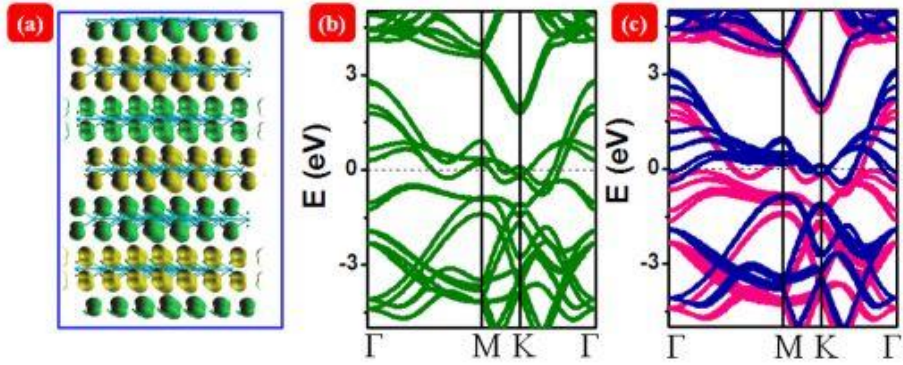

Figure 26 (a) Spin density in the 3R−MoN$_2$ bulk (green colour for spin-up electrons, yellow colour for spin-down electrons). (b) Band structures of the 3R−MoN$_2$ bulk. (c) Spin-polarized band structure of bulk 3R−MoN$_2$ (red lines for spin-up, blue lines for spin-down). Reproduced with permission [41]. Copyright (2015), American Chemical Society.

The MoN$_2$ monolayer has been predicted to be a SGS material, as shown in Figure 26. The MoN$_2$ monolayer has ferromagnetism and the Curie temperature is up to 420 K.[41] Moreover, the local magnetic moments are from the N$p$ orbitals. The Curie temperature is quite higher than that in other 2D atomic crystals with flat surfaces. The magnetism is intrinsic, and external doping and structural modification are not required for the generation of the magnetism.

3.3.3 Na$_2$C monolayer

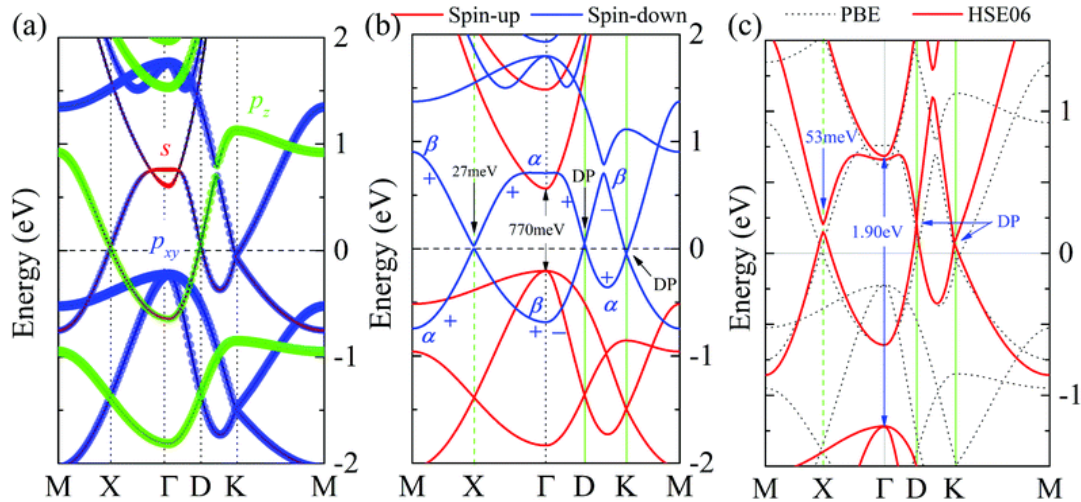

Figure 27. (a) Orbital-related band structures in the Na$_2$C without SOC. (b) Spin-related band structure in the Na$_2$C without SOC. (c) Band structures in Na$_2$C with SOC. Reproduced with permission[42]. Copyright (2018), Royal Society of Chemistry.

In addition, the Na$_2$C monolayer has been proposed as a $p$-state half-metal with several Dirac cones.[42] It has several SGS features, as shown in the Figure 27. It owns symmetry protected and spin-polarized Dirac cones. It also has gapped Dirac nodal line. It has a high Fermi velocity and ferromagnetic ground state. The Curie temperature is up to 382 K. The ferromagnetism results from unpaired 2$p$ electrons in the C. The 2$p$ magnetism is from the superexchange mechanism between C$^{2-}$ and Na$^+$.

4. Parabolic-dispersion SGS materials

Some oxides monolayers, such as $VO_2$ and MnO with a honeycomb and square lattice, have SGS band structures and parabolic dispersion. Many oxide materials have honeycomb lattices, such as the ZnO, which has a hexagonal structure. The MnO monolayer with saturation of H atoms also holds the SGS feature with parabolic dispersion.[43] The same type of SGS with parabolic dispersion has also been found in ferromagnetic monolayer CoO.

5. Two-dimensional (2D) SGS materials

Many 2D SGS materials have been proposed as SGS materials in the last few decades. They include: (1) Armchair graphene nanoribbons combined with $CH_2$, O and NH radical groups;[44] (2) Transformed zigzag graphene nanoribbons through *n*-doping defects;[45] (3) Tuned graphene nanoribbons with sawtooth edges;[46] (4) Fe and Cr co-doping in boron nitride sheets;[47] (5) Ferromagnetic $HgCr_2Se_4$ compound under a pressure;[48] (6) Boron nitride nanoribbons with some B or N vacancies;[49] (7) Single atomic Fe-Fe chains with zero gap.[50] (8) 2D pristine $MnX_3$ (X = F, Cl, Br, I) is a family of SGS materials with a gap in one spin channel and a Dirac cone in the other.[38] Besides, the ferromagnetic monolayer $VO_2$ with a square lattice has the linear spin-polarized gapless band structure.[3] The monolayer $VO_2$ owns linear dispersion around Fermi level and spin direction for both the CB and VB is same.

One of the well-known SGS materials is cobalt-doped $PdPbO_2$ films.[51] This compound has a weakly temperature-dependent resistivity. Based on band structure calculations, cobalt-doped $PdPbO_2$ owns a zero-gap feature with spin-up band touching spin-down band near Fermi level. Colossal electroresistance (ER) and giant MR were found in the doped $PbPdO_2$ compound.[51] Below the metal-insulator transition, the electrical current strongly supresses the resistivity and the ER values is up to $10^7$ Ω. This prediction was verified through the study of electronic structure in the doped $PbPdO_2$ films using photoemission spectroscopy (PES) as well as soft X-ray absorption spectroscopy (XAS).[52] Mn-doped ZnTe has a DOS similar to a SGS material.[53] In this material, the maximum of the spin-up VB and the minimum of spin-down CB meet at Fermi level. The VB is completely filled and the spin-down CB is empty. Hence, charge carriers vanish in the compound, which leads to antiferromagnetic exchange interactions.

6. Summary and outlook of the SGS materials

| Types of SGS materials | Typical materials | Physical features |
|---|---|---|
| Heusler SGS materials | MnCoAl bulks | High Curie temperature |
| Dirac SGS materials | $CrO_2/TiO_2$ heterostructures | High mobility |
| Parabolic SGS materials | MnO and CoO monolayers | Honeycomb lattice |
| 2D SGS materials | $PdPbO_2$ films | Colossal electroresistance |

Table 1. Summary of the main types of the SGS materials.

The SGS materials are one type of new materials with fully spin polarised electrons or holes. The band structures of SGSs could be Dirac type or parabolic type. SGS materials are all ferromagnetic materials with high Curie temperatures. Dirac type SGS materials also have relatively high mobility. The theoretical predictions are mostly based on first principle calculations. A number of materials, including both monolayer materials and bulk materials, have been predicted as SGS materials (see Table 1). They hold great potential for high speed and low-energy-consumption spintronics and electronics. There are still few SGS materials that have been experimentally realized, and there is a great potential to find new materials. The reason is that most predicted SGS materials are monolayer materials, which are hard to synthesize. Also, some monolayer materials are not stable in the ambient environment. New nanotechnology is needed for further fabricating monolayer SGS materials. New devices can then be explored and fabricated after experimental realization of monolayer SGS materials.

Acknowledgements: We acknowledge the funding support from the ARC through the Centre of Excellence FLEET project. We thanks Mrs Tania Silver for contribution in English polishing.

Conflict of Interest: The authors have no Conflict of Interest.